**Dynamical Geochemistry: Mantle dynamics and its role in the formation of geochemical heterogeneity**

Peter E. van Keken[1,*], Catherine Chauvel[2], and Christopher J. Ballentine[3]

[1]Earth and Planets Laboratory, Carnegie Institution for Science, 5241 Broad Branch NW, Washington DC 20015, USA. pvankeken@carnegiescience.edu

[2] Université Paris Cité,, Institut de Physique du Globe de Paris, 1, rue Jussiee, 75005 Paris, France. chauvel@ipgp.fr

[3] Department of Earth Sciences, Oxford University, South Parks Road, Oxford OX1 3AN, United Kingdom. chris.ballentine@earth.ox.ac.uk

[*] Corresponding author

## Abstract

Chemical geodynamics is a term coined nearly forty years ago to highlight the important link between Earth's geochemical evolution and plate tectonics & mantle convection. Significant progress in our understanding of this connection has taken place since then through advances in the analytical precision of geochemical measurements; dramatically improved geophysical imaging techniques; application of novel isotope systems; and great advances in computational power. The latter especially has improved geodynamical models and data interpretation techniques. We provide a review of these advances and their impact on chemical geodynamics, or perhaps, dynamical geochemistry. To focus this review we will address primarily the role of whole mantle convection and oceanic crust formation and recycling together with an update on our understanding of noble gas systematics.

## Glossary and nomenclature

MORB(s) – Mid-Oceanic Ridge Basalt(s)
OIB(s) – Ocean Island Basalt(s)
HW82 – Hofmann and White (1980, 1982)
EM1 – Enriched Mantle component 1
EM2 – Enriched Mantle component 2
HIMU – OIB component with high original $\mu$=U/Pb
LLSVPs – Large Low Shear Velocity Provinces
ULVZ – Ultra Low Velocity Zone
BSE – Bulk Silicate Earth
CHUR – CHondritic Uniform Reservoir

# 1. Introduction

Early observations of heterogeneity in lithophile isotopes measured in mid-oceanic ridge basalts (MORBs) and ocean-island basalts (OIBs) led to the conceptual idea of a layered mantle system. In this model, the upper mantle is depleted by prior extraction of the continental crust and OIBs are explained by mixing between the depleted mantle and a lower mantle that has been largely isolated from plate tectonics and mantle convection (e.g., DePaolo and Wasserburg, 1976; O'Nions et al., 1979; Allègre et al., 1979). Noble gas systematics appeared to provide strong support for such a layered system. For example, the $^3$He/$^4$He measurements in OIBs are wide-ranging whereas the ratio is relatively uniform in MORBs (e.g., Barfod et al., 1999; Figure 1a). In addition, the fundamental observation that ~50% of $^{40}$Ar that has been produced from $^{40}$K decay resides presently in the atmosphere and upper mantle concentrations are low, was fully consistent with a lower mantle that is a largely undegassed reservoir (Hart et al., 1979; Allègre et al., 1996).

Lithophile isotope ratios in OIBs and MORBs suggest mixing trends between various components such as the two Enriched Mantle components (EM1 and EM2), HIMU (which is formed from material that had high original $\mu$=U/Pb), and MORB. From some perspectives these mixing trends seem to converge (Figure 1b) on a FOcal ZOne (FOZO; Hauri et al., 1994) (alternatively called 'C' for Common component or PREMA for PREvalent MAntle; Zindler and Hart, 1986; Hanan and Graham, 1996) which paradoxically has relatively high $^3$He/$^4$He but appears depleted in other regards (see discussion in van Keken et al., 2001). It should be noted that the layered mantle hypothesis, in which the upper mantle is primarily composed of well mixed MORB-source material with the lower mantle supplying the poorly mixed OIB components, is at least somewhat contradicted by global observations of isotopic heterogeneity. Lead isotope systematics, for example, show an OIB spread of data that is somewhat more diverse, but largely overlaps with the MORB data (compare Figure 1c and 1d which are redrawn on the same scale from Hofmann, 2014).

Hofmann and White (1980, 1982; hereafter called HW82) provided a striking counterview to the layered convection hypothesis by suggesting that the long term formation of oceanic crust and its recycling and sequestration in the deeper mantle could explain the observed isotopic heterogeneity (Figure 2a). It is useful to quote the abstract of Hofmann and White (1982) in its entirety:

> We propose the following model for the origin of "hot-spot" volcanism: Oceanic crust is returned to the mantle during subduction. It is separated from the surrounding peridotite, it sinks into the deeper mantle and accumulates at some level of density compensation, possibly the core-mantle boundary. The accumulated layer locally reaches thicknesses exceeding 100 km. Eventually, it becomes unstable as a consequence of internal heating, and the resulting diapirs become the source plumes of ocean island basalts (OIB) and other hot-spot *volcanism. This mechanism may trigger upper mantle convection as suggested by Morgan [1]. Our model provides possible explanations for (1) the high trace element concentrations of OIB, (2) the seemingly contradictory isotopic evidence for both enrichment and depletion of magmaphile elements in OIB sources, (3) the phenomenon of apparent mantle isochrons in oceanic basalts, and (4) the apparently episodic nature of continental igneous activity. The model can be tested further, as more knowledge accumulates about the actual bulk composition of the oceanic crust, including its alteration products and sediment cover.*

In this review we have two main goals. The first is to address how the HW82 hypothesis has been tested and expanded upon by geochemical, geophysical, and geodynamical studies. The second is to provide an update on our understanding of noble gas systematics and how they constrain the evolution of the Earth's interior.

This review chapter is not intended to capture all advances in "chemical geodynamics" (Zindler and Hart, 1984) that have been made since the publication of HW82. Rather, it will provide an exploration of how well main hypothesis in HW82 has stood the test of time amidst new geochemical techniques & observations and abundant geophysical evidence for heterogeneity in the lowermost mantle. Previous reviews that further explored chemical geodynamics in a broader sense are provided by van Keken et al. (2002), G.F. Davies (2011), and Tackley (2015). Basic descriptions of mixing processes that lead to the formation and destruction of geochemical heterogeneities are in Kellogg (1992), van Keken et al. (2003), and van Keken (2013). A review linking mantle mixing with possible seismic observations is in Stixrude and Lithgow-Bertelloni (2012). Updated overviews of the geochemical observations in MORBs and OIBs are in Stracke et al. (2005), Hofmann (2014), Hanyu and Chen (2021) and Weis et al. (2023). Finally, overviews of the nature of LLSVPs and their geophysical relationships to oceanic crust recycling are provided by McNamara (2019) and M. Li (2021).

## 2. Geophysical implications of long-term oceanic crust recycling.

### 2.1 Recycling of oceanic lithosphere and the formation of LLSVPs

We will take the viewpoint that plate tectonics has been in operation for most of Earth's history. The long-term formation of chemically heterogeneous oceanic lithosphere at mid-oceanic ridges and its recycling at subduction zones may provide a natural explanation of the formation of the Large Low Shear Velocity Provinces (LLSVPs) that are seen at the base of the Earth's mantle. A simple calculation adds quantitative support to this hypothesis. Reymer and Schubert (1984) suggest 675 $km^3$/km/Myr of oceanic crust is being subducted. With a total length of 37,000 km of destructive plate margins and an age of 4.5 Gyr for the Earth we find that $1.1 \times 10^{11}$ $km^3$ of crust has been recycled. This is 13% of the volume of the mantle. The total volume of the LLSVPs has recently been estimated to be about 8% of that of the mantle (Cottaar and Lekic, 2016). Since the LLSVPs are not composed of pure oceanic crust but of a mixture of oceanic crust, harzburgite, and ambient mantle there appears to be a more than sufficient amount of oceanic crust has been recycled to allow for gradual but steady re-entrainment of oceanic crust from the piles. A corollary of this back-of-the-envelope computation is that with an average melt fraction of 10-20% the oceanic crust production estimate suggests between 65–130% of the mantle has been processed at a mid-oceanic ridge. Since parts of the mantle will be processed more than once, it seems very unlikely that the whole mantle will have lost its primitive signal by surface melting.

---

[1] Morgan W.J, 1971. Convection plumes in the lower mantle. Nature 342, 42-43.

A first convincing dynamical model that shows the formation of LLSVP-like structures due to recycling of oceanic lithosphere with dense crust was provided by Christensen and Hofmann (1994; Figure 2d). Computational limitations at the time restricted their models to lower convective vigor than present day's but a scaling argument allowed them to extrapolate their results to realistic convective vigor. This enabled them to determine that the accumulated oceanic crust imparted a chemical signature on the model that is similar to that observed in Nd-Pb and Pb-Pb data with a pseudo-isochron of 2.1 Gyr. Brandenburg and van Keken (2007) confirmed these models and were able to use Earth-like convective vigor that led to similar findings. Significant other work that shows the possibility of long-term oceanic lithosphere recycling causing LLSVPs is provided in G.F. Davies (2002), Brandenburg et al. (2008; Figure 2b), Nakagawa et al. (2012; Figure 2c), Mulyukova et al. (2015), Huang et al. (2020), Jones et al. (2021; Figure 2e), M. Li and McNamara (2022), and Panton et al. (2023).

## 2.2 Seismological observations

### 2.2.1 Tomography

Indications for large scale compositional heterogeneity in the deep mantle are seen in the earliest tomographic models that used low-order spherical harmonics (e.g., Dziewonski et al., 1977; Dziewonski, 1984; Woodhouse and Dziewonski, 1989). Significant improvements in seismic tomography have allowed convincing imaging of two nearly anti-podal LLSVPs which are dominant in all modern tomographic models (see reviews in Garnero et al., 2016, and Cottaar and Lekic, 2016). A common interpretation of these regions is that they are both warmer and denser than ambient mantle, forming long-lived thermochemical piles that are formed either from primordial heterogeneity that has not been fully entrained by mantle convection, subducted oceanic lithosphere, where the eclogitic nature of the oceanic crust causes the high density, or a combination thereof. Mantle convection modeling lends support to such an interpretation where continuous thermochemical piles are kept in shape by surrounding downwellings (see discussion in Garnero et al., 2016). An interesting contrasting interpretation from the full waveform modeling based tomographic model SEMUCB-WM1 (French and Romanowicz, 2014) is that the features that make up the LLSVPs appear to be clusters of thermochemical plumes, rather than continuous piles (Davaille and Romanowicz, 2020). Of course, the chemically distinct material in this interpretation could remain composed of recycled oceanic lithosphere with denser oceanic crust.

Seismic tomography approaches can be combined or augmented with other geophysical constraints to confirm the high density of the LLSVPs. For example, Ishii and Tromp (1999) showed by incorporating free-air gravity measurements that the density is higher in the regions of lower P- and S-wave velocity, suggesting a thermochemical origin for the LLSVPs. Employing the sensitivity of Earth's diurnal tides to the density distribution in the mantle, Lau et al. (2017) showed that the LLSVPs have an average density contrast with the ambient mantle of 0.5-0.8%.

It has become possible to use sensitivity kernels developed for some tomographic models, such as S20RTS (Ritsema et al., 1999) and S40RTS (Ritsema et al., 2011), to explicitly test whether a certain dynamical model, after mapping temperature and composition to seismic velocities, can reproduce the features seen in the tomographic models. An early example is in Bull et al. (2009) who argued that thermochemical piles fit the observed characteristics better than plume clusters. A counterpoint was provided by Schuberth et al. (2009) and D.R. Davies et al. (2012) who argued for a purely thermal origin of the LLSVPs. Recently, Jones et al. (2020) showed that the long-term oceanic crust recycling as in the models of Brandenburg et al. (2008) could reasonably explain the imaged velocity contrast and extent of the two LLSVPs in S40RTS if the density contrast between oceanic crust and ambient mantle was 3.8% or higher. Similarly, the spectral content of the filtered geodynamical models appears similar to that in S40RTS for higher density contrasts, but not when the density contrast between oceanic crust and mantle is low (see Figure 5 in Jones et al., 2020). Since the thermochemical piles in these models are mixtures of oceanic crust, depleted mantle, and ambient mantle, the density contrast between crust and mantle should not be taken to be representative of the density contrast between LLSVP and ambient mantle – clearly this contrast will be significantly lower.

## 2.2.2 Scattering

High-frequency waveform analysis allows for the determination of compositional heterogeneity in the deep mantle at much higher resolution than what seismic tomography can provide (see review in Kaneshima, 2016). Scattered wave energy in the lower mantle generally is interpreted as being caused by subducted oceanic crust. For example, Kaneshima and Helffrich (2009) observed clusters of scatterers below Pacific subduction zones at depths of 1100–1800 km that could be explained by seismically distinct basalt layers in folds with spacing of 100–200 km, which is similar to what we expect for coherent slabs in the early stages of folding. Other recent evidence from scattering for the presence of coherent oceanic crust in the lower mantle below subduction zones is provided by, e.g., He and Zheng (2018), Zhang et al. (2020), and Ritsema et al. (2020).

Haugland et al. (2018) provided a direct test whether long-term recycled oceanic lithosphere characterized by a denser basalt component could explain PKIKP precursors that are related to scattering of PKP waves by heterogeneity in the deep mantle. They used the models of Brandenburg et al. (2008) to predict P-wave velocity perturbation and modeled seismic wave propagation using the axisymmetric waveform modeling method AxiSEM (Nissen-Meyer et al., 2014). They showed that models where the oceanic crust had the same density as the surrounding mantle could not explain the PKIKP precursors but that models where the oceanic crust had an excess density could. They only investigated the latter in a model where the oceanic crust had an excess density with respect to that of the depleted harzburgitic component of 7%. This translates to an excess density of the crust with respect to 'normal' peridotitic mantle of 4.3% (see discussion in Jones et al., 2020). Frost et al. (2017, 2018) also studied scattering of PKP waves and showed that scattering intensity increased in the lowermost 1000 km of the mantle and that the scattering intensity was correlated with regions of low seismic velocity with some stronger scattering near the edges of the LLSVPs, which they explained with a model where oceanic crust accumulates on the sides of piles of pre-existing heterogeneity. Hiemer and Thomas (2022) observed scattering at the bottom 200–400 km of the mantle below the mid-Atlantic and determined from comparison with synthetic models that the scatterers had a correlation length of 10 km and velocity contrast of 5% which could be caused by subducted oceanic crust, although their work could not distinguish this from other potential features in D" such as Ultra Low Velocity Zones (ULVZs) with an iron-rich post-perovskite component (see Ma and Thomas, 2020 and references therein) or melt pockets (Hedlin and Shearer, 2000). Kaneshima (2023) demonstrated the presence of scatterers in the transition zone and lower mantle below the Samoan hotspot that is interpreted as caused by basaltic materials entrained in the Samoan plume rising from the Pacific LLSVP.

## 2.3 Entrainment of dense lower mantle piles by plumes

The HW82 hypothesis relies on the return of the matured oceanic crust to the Earth's surface, which requires either slow entrainment by large scale mantle convection or direct entrainment in mantle plumes rising from the LLSVPs. Since plumes are generally associated with the transport of heat and geochemical heterogeneity from the deep Earth, the latter mechanism may be the most direct and efficient, but this also requires that thermal plumes are capable of entraining the dense material. Significant work in the last two decades has studied the interaction of a dense layer at the base of the mantle with thermal plumes rising from it. This has demonstrated that plumes are strongly modified from the classical mantle plume shapes seen in early, and inspiring, laboratory experiments (e.g., Griffiths and Campbell, 1990). Farnetani and Samuel (2005) suggested we should abandon the thermal plume paradigm in favor of seeing plumes as complex geometrical entities. Lin and van Keken (2005; 2006a; 2006b) showed that the interaction between thermal and (negative) compositional buoyancy leads to time-dependent effects in axisymmetric spherical plumes that could explain temporal variations in ocean island chain volcanism but also that plumes might stall at depth due to an effective neutral buoyancy, a mechanism also demonstrated by Samuel and Bercovici (2006). Dannberg and Sobolev (2015) showed how the effect of entrainment of eclogite could cause large plumes that would not cause significant uplift, possibly explaining the formation of large igneous provinces, a mechanism demonstrated earlier by Lin and van Keken (2006b).

      The numerical studies above are complemented by laboratory studies. While perhaps not every Earth-like scenario can be simulated in a laboratory, these models certainly have the advantage that they, by their very nature, accurately satisfy the governing principles of conservation of mass, momentum, and energy. These same principles form the basis of partial differential equations that are solved using approximate methods in numerical studies. Figure 3a shows the now classic example of rising thermochemical plumes from Kumagai et al. (2008).

The orange layer has a higher density and temperature is indicated by the lighter contour lines (that are formed by crystals in the fluid that turn opaque over a narrow temperature range). The relative density contrast between the dense layer and the thermal effect is indicated by the buoyancy number B which increases from left to right in Figure 3a. The buoyancy number is the ratio of the estimated chemical buoyancy divided by the estimated thermal buoyancy, that is, $B=\Delta\rho/(\rho\,\alpha\,\Delta T)$ where $\Delta\rho$ is the compositional density contrast, $\rho$ is the density of the ambient fluid, $\alpha$ is the thermal expansivity, and $\Delta T$ is the temperature contrast across the bottom boundary layer. This provides a clear demonstration that the plumes rise easily with a compositional component at low B<0.3, that plumes can stagnate at intermediate B (0.3–0.9), and that plumes rise again as thermal plumes leaving the dense layer behind at high B > 1. Given that the estimated chemical excess density from seismological observations and from numerical experiments is ~1%, it follows (with the reasonable values $\alpha=2\times10^{-5}$/K and $\Delta T=1500$ K) that B~0.33, suggest plume entrainment of the dense material is moderately efficient.

Some detailed comparisons between laboratory and numerical experiments exist that give us confidence that these approaches provide quantitative information about thermochemical plumes rising through the Earth's mantle. Early comparisons between laboratory and numerical experiments of thermal plumes showed good general agreement if the right laboratory conditions were taken into account (van Keken, 1997). A detailed study comparing thermal plumes with finite element methods using an axisymmetric cylinder geometry was provided by Vatteville et al. (2009) showing excellent agreement in the evolution of the temperature and velocity fields. Figure 3b shows one example that is similar to the first frame in Figure 3a. Minor differences in the detailed temporal evolution that were left unexplained were resolved by modeling of the plumes in the actual 3D Cartesian box geometry of the laboratory experiment (D.R. Davies et al., 2011) showing that accurate implementation of the boundary conditions in Stokes flow remains important. To our knowledge there are no such detailed comparisons available for thermochemical plume simulations, but a comparison between the existing models of Kumagai et al. (2008) and Lin and van Keken (2006b) provides at least good quantitative agreement (compare Figure 3 frames c1 and c2 with frames of the Kumagai experiments that are obtained at similar buoyancy number as those in the numerical experiments.)

## 3. Geochemical consequences: the lithophile element/isotope perspective

When Hofmann and White suggested in 1982 that recycled oceanic crust played a crucial role in the origin of plume generated magmas, their main argument was that ocean island basalts are rich in very incompatible elements and such enrichments cannot be produced by melting of a normal peridotite source. Indeed, to reproduce the light rare earth content of OIB requires very small and in some cases impossibly small degrees of melting (Gast, 1968) if the source is assumed to be the same as that of mid-oceanic ridge basalts. At that time, the number of radiogenic isotopic data for OIB was very small and most of them had values intermediate between a depleted mantle, MORB source, and a non-differentiated reservoir often called

CHUR (CHondritic Uniform Reservoir) or BSE (Bulk Solid Earth). However, few ocean islands had Sr and Nd isotopes that required the presence in their source of material more enriched than BSE, an observation that weakened the generally accepted model of mixed MORB source (upper mantle) and BSE source (lower mantle) for plume magmatism, and provided support for HW82.

The evidence for the presence of recycled oceanic material in the mantle has now grown drastically and is not necessarily limited to the source of ocean island basalts. This includes evidence from the major and trace element concentrations, evidence from the radiogenic isotopes but also evidence from traditional and non-traditional stable isotopes.

From a major and minor elements point of view, several authors (for example, Hauri, 1996; Hirschmann and Stolper, 1996; Kogiso et al., 1997) demonstrated early on that the presence of recycled oceanic crust could explain the main geochemical and petrological features of intra-plate volcanism but also that of mid-oceanic ridges. However, a big step forward happened when Sobolev et al. (2005) highlighted that the high nickel and silica contents of Hawaiian lavas could not result from simple melting of a peridotite source. Instead, they suggested that in a rising plume a recycled oceanic crust present as eclogite reacts with the surrounding peridotite to produce a pyroxenite that in turn melts when approaching the lithosphere under Hawaii. In the case of Hawaii, the proportion of recycled crust in the plume source was estimated at about 20% but in other cases, the proportion of recycled crust can be much lower. For example, it is estimated to be ~10% for Iceland (Sobolev et al., 2007) and ~5% for Gambier in the Pitcairn chain (Delavault et al., 2015). While such interpretation is generally well accepted by the community, alternative views have also been provided (see, for example, Herzberg et al., 2014, or Weiss et al., 2016).

The clearest geochemical indices of the presence of recycled oceanic crust (which generally includes a sedimentary cover) comes from the radiogenic isotopes because sediments have both isotopic compositions and trace element contents that differ drastically from those of the mantle. Indeed, the presence of few percent of sediment in the source of an OIB has a significant effect on the isotopic composition of the lavas. The best example was published by M.G. Jackson et al. (2007; Figure 4a). They demonstrated that the presence of 6% sediment in the mantle source of some Samoa lavas shifted their Sr isotopic composition from 0.704 to extreme values (~0.720) not measured in any other OIB. However, most EM2 basalts have significantly lower radiogenic Sr isotope ratios than those detected in Samoan lavas, but their Sr isotopic composition still requires a sediment contribution up to 2%. Altogether, the community agrees that most OIBs with radiogenic Sr coupled to unradiogenic Nd and Hf must come from a source containing some recycled sedimentary material (for a review see Stracke, 2012). Other evidence for the presence of recycled sediments comes from the relationship between Nd and Hf isotopes in ocean islands (see figure 4c and Chauvel et al., 2008) because mixing recycled basaltic crust and depleted mantle creates an array that is not consistent with the known OIB array. A small but significant proportion of sediment is required to reproduce the OIB mantle array (Figure 4b).

Evidence for the presence of recycled basaltic crust without sediment is only found in the so-called HIMU ocean islands (Figure 4b). Their very radiogenic Pb isotopes require a large increase of their U/Pb and Th/Pb ratios about 2 Gyr ago, an increase that is most easily explained by the loss of Pb during dehydration of the basaltic crust in the mantle wedge of subduction zones. Basaltic crust recycled more recently than 2 Gyr ago is more difficult to identify from a radiogenic isotopes point of view because there is not enough time for the change in parent-daughter ratio to have an impact on the daughter isotopic ratio (Figure 4b). The consequence is that any recycled basaltic crust younger than 1.5 Gyr ago has Pb isotopic ratios that are subdued and difficult to identify and the presence of significant amounts of such material in the mantle cannot be pinned down from a radiogenic isotope perspective.

From a trace elements point of view, evidence for the presence of recycled crust in the source of mantle magmas includes of course the enrichment in incompatible trace elements of OIB, enrichment impossible to produce by simple melting of the mantle. Other evidence is more subtle and highlights the involvement of sedimentary material associated to the subducted basaltic crust and present in the material recycled in the mantle. The clearest cases are the deviation from mantle values of ratios of cerium to lead (Ce/Pb) and niobium to uranium (Nb/U). Indeed, Hofmann et al. (1986) demonstrated that MORB and OIB shared constant Ce/Pb (25±5) and Nb/U (47±10) but these 'canonical' values are very different from both BSE (Ce/Pb=9 and Nb/U=30) and Continental Crust (Ce/Pb=4 and Nb/U=10). Such constancy implies that Ce and Pb on one hand, and Nb and U on the other, have similar incompatibilities during mantle melting. Any deviation from the canonical values requires that material with an exotic origin is present in the source of the volcanics. If OIB volcanics have a Ce/Pb and/or Nb/U ratio lower than the canonical values, it potentially suggests that continental material contributes to the source of the lavas. Qualitatively, material with a primitive mantle origin would also lower the two ratios but mass balance calculations show that the proportion of BSE required to modify Ce/Pb or Nb/U is much larger than if the material is of crustal origin. Few OIBs have lower than average Ce/Pb and Nb/U and those include several islands in the Polynesian archipelagos (Society: White and Duncan, 1996 and Cordier et al., 2021; Marquesas: Chauvel et al. 2012; Pitcairn-Gambier: Cordier et al., 2021) but also islands in the Samoa chain (M.G. Jackson et al., 2007). Among mid-oceanic ridge basalts, the Central Indian ridge also stands as an exception with its slightly lower than average canonical ratios (Rehkämper and Hofmann, 1997) suggesting the presence of poorly mixed recycled sediments in that part of the upper mantle. At the other end of the spectrum are found few islands with higher than average ratios. We can cite here St. Helena in the South Atlantic, several islands along the Austral and Pitcairn chains, and lavas from the Canaries.

Traditional and non-traditional stable isotopes also prove useful in tracing in plume sources the presence of material that once was at the surface of the Earth. The most convincing evidence comes from sulfur isotopic analyses of some ocean islands. Indeed, prior to the Great Oxidation Event 2.4 Gyr ago, the atmosphere was not as oxygen rich as today and material formed at the surface of the Earth had sulfur isotopic compositions that deviated from a mass fractionation line (Farquhar et al., 2007). Cabral et al. (2013) found that such mass-independent fractionation occurred in sulfides present in lavas from Mangaia island. Delavault et al. (2016) reported even larger anomalies in samples from Pitcairn seamounts (Figure 5). The presence of such anomalies in present-day volcanics requires the existence in their source of material that not only was present at the surface of the Earth at some stage but also that this material is older than 2.4 Gyr ago.

Other potential indices of the presence of surface material in the source of OIB can be found in the isotopic compositions of lithium or thallium, two elements that are significantly fractionated in low temperature environments (see, e.g., Penniston-Dorland et al., 2017, Rehkämper et al., 2002 and Nielsen et al., 2017) but not in mantle conditions. For example, lithium isotopic heterogeneities have been reported for islands such as Hawaii, Rurutu, St. Helena, and the Azores (Chan and Frey, 2003; Vlastélic et al., 2009; Krienitz et al., 2012; L.N. Harrison et al. 2015) and large and significant variations of thallium isotopes have been reported by several groups (Blusztajn et al., 2018; Brett et al., 2021; Williamson et al., 2021). Finally, elements such as magnesium or calcium are not diagnostic because the isotopic variability is small among OIB but the fact that their isotopic ratios are not constant suggests that heterogeneities produced at the Earth's surface might be integrated in their source deep in the mantle (see Teng, 2017, and review by Soderman et al., 2022).

## 4. Geochemical consequences: noble gases and volatiles

### 4.1 Foundational context

The inert, or noble, gases have provided an observational cornerstone in our determination of the geochemical origins and geodynamic evolution of the Earth's silicate mantle (see reviews by, e.g., Porcelli and Ballentine, 2002; Ozima and Podosek, 2002; Moreira, 2013; Péron et al., 2018; Mukhopadhyay and Parai, 2019). Noble gases in the Earth's mantle can be grouped into four types: i) those trapped during the Earth's accretion; ii) those formed by the decay of extinct radionuclides ($^{244}$Pu, $^{129}$I) in the early history of the Earth ; iii) gases generated by the decay and interaction from extant radionuclides over Earth history (e.g. U, Th, K); and iv) atmospheric noble gases recycled into the mantle. Each has a distinct and resolvable isotopic signature within the 23 stable isotopes that form the noble gas system. The varying proportions of these different noble gas components found in different mantle samples and their geological and tectonic contexts provide a unique insight into the dynamics of the mantle that have preserved these volatile records and variably mixed them over Earth history.

The ground breaking recognition of a primordial $^{3}$He signature in volcanic gases and seawater (Mamyrin et al., 1969; Clarke et al., 1969) identified this as a volatile gas that was incorporated into the Earth during its accretion. Since helium is lost by Jean's escape (Jeans, 1925), the Earth's atmosphere has a distinct $^{3}$He/$^{4}$He isotopic composition (1.4×10$^{-6}$ = 'R$_a$') and low concentration (5.4 ppm) compared to other volatiles. This results in both low contamination of helium in mantle samples by seawater or atmosphere exposure during eruption or degassing, and negligible recycling back into the mantle; simplifying its interpretation as a 'primordial' gas trapped during the accretion of Earth.

Primordial $^3$He together with $^4$He that is generated by the decay of U and Th means that the $^3$He/$^4$He isotopic composition of the mantle evolves as a function of time through the evolution of $^3$He/(U+Th) and convective mixing. Mantle $^3$He/$^4$He and its variance provides an unprecedented insight into the interaction of the chemical reservoirs in the Earth's mantle. Indeed, the observation that plumes forming oceanic islands sample a portion of the mantle with a higher $^3$He/$^4$He than the mantle supplying mid-oceanic ridges demands long-lived chemical heterogeneity within the mantle system (Kurz et al., 1982; Allègre et al., 1987)

With mantle-derived helium readily resolved in ocean water, gas residence and the distribution of $^3$He in the oceans provide an estimate of the present day helium ($^3$He and $^4$He) degassing flux (1060 mol $^3$He per year; Lupton & Craig, 1975; Farley et al.,1995). Combined with the ocean crust formation rate (20 km$^3$/yr) and average partial melting forming the crust (10%) the $^3$He flux in turn provided an estimate of the upper mantle $^3$He concentration (1.18×10$^9$ atoms $^3$He per gram); at the time, arguably the most robust estimate of the concentration of any mantle volatile (Kellogg and Wasserburg, 1990). Identification of the ratio of $^3$He to other volatiles (such as the heavier noble gases, carbon, water, and nitrogen) in mantle samples thus provides a primary reference for determining their concentration, origin, and evolution in the mantle (Craig and Lupton, 1976; Kurz et al., 1982; Allègre et al., 1983; 1987; Marty and Jambon, 1987; Javoy and Pineau, 1991; Trull et al., 1993; Marty, 1995).

Unlike $^4$He, $^{40}$Ar accumulates in the atmosphere. Because it is generated by β$^+$ decay of $^{40}$K it provided an early insight into the volume of the mantle that must have been degassed and was identified as a direct measure of mantle convection dynamics (Hart et al., 1979). Refinement of the BSE K content led to an estimate of mantle degassing efficiency of circa 50% (Allègre et al., 1996). This agrees with the outgassing efficiency derived using the excess radiogenic $^{136}$Xe observed in the atmosphere, resulting from $^{144}$Pu and U decay (circa 40% degassing; Porcelli and Ballentine 2002). Since U and Pu are refractory lithophile elements these are independent of concerns about the BSE estimate for the more volatile K (Lassiter, 2004).

It follows that the $^{40}$Ar generated by the BSE K must be the sum of that in the mantle and atmosphere. The concentration of $^{40}$Ar in the mantle, calculated from the $^3$He and $^3$He/$^{40}$Ar* (where $^{40}$Ar* is the $^{40}$Ar in excess of that from atmospheric contributions), is low and does not fulfil this simple mass balance if this concentration is extrapolated to the whole silicate mantle. This provides the basis for conceptual models requiring an unsampled region of the mantle to contain significantly higher concentrations of $^{40}$Ar to make up the difference. This region or domain is often equated with the source of high $^3$He/$^4$He seen in OIBs (Kurz et al., 1982; Allègre et al., 1996). Further key observations that provide the foundations on which, together, the structure and dynamics of the Earth's mantle start to be defined include the observed mantle heat flow. While consistent with the whole mantle U+Th, this is higher than the concomitant flux of $^4$He accounted for by the U+Th in the upper mantle (depths less than 660 km) alone. This led to the proposal that the 660 km boundary layer in the Earth's mantle provided a barrier through which heat but not He could efficiently escape (O'Nions and Oxburgh, 1983). In contrast, the calculated concentration of $^3$He relative to U+Th in the mantle underlying mid ocean ridges if the system were a simple closed 'box' (Jochum et al., 1983) predicts a more radiogenic helium $^3$He/$^4$He signature (2.6 R$_a$) than that observed (8 R$_a$). This difference between model and observation required a helium flux from deeper in the mantle system (below 660 km) to buffer the $^3$He/$^4$He of the upper mantle (Allègre et al., 1983; 1987). Such a flux is compatible with the 'heat-helium paradox' only if the flux from the deep to shallow mantle is high in $^3$He and low in $^4$He – which is consistent with the observed high $^3$He/$^4$He in OIB (Figure 1a).

Models that assume the flux of gas from the different reservoirs has reached steady state were first built to account for He (Kellogg and Wasserburg, 1990), He, Ne, and Ar (O'Nions and Tolstikhin, 1994) and then all noble gases (Porcelli and Wasserburg, 1995a; 1995b), with the latter notably identifying recycling of heavy noble gases back into the mantle necessary for the Xe mass balance. These papers were foundational in showing that the abundance and isotopic composition of the noble gases in the upper mantle (above 660 km) and atmosphere could be broadly described by the same layered mantle construct used to account for the radioelement isotopic evolution of the mantle and crust (e.g., DePaolo and Wasserburg, 1976; O'Nions et al., 1979; Allègre et al., 1979). The noble gas models nevertheless also placed firm constraints on the dynamics between the upper and lower mantle (limited mass exchange), the noble gas abundance and isotopic composition of the lower mantle (higher noble gas concentrations and more primitive and therefore higher $^3$He/$^4$He and lower $^{40}$Ar/$^{36}$Ar when compared with the upper mantle) and relationship between the lower mantle and upper mantle volatiles (the deep mantle gas fluxes to the upper mantle). With new observations from seismic tomography, dynamical modeling, and deep mantle geochemistry these layered mantle constructs are no longer tenable but the requirement to account for the same observational foundations remains.

The dynamical models of Brandenburg et al. (2008) point to a resolution of the missing Ar problem. Even while these models have Earth-like convective vigor as measured by plate velocities and surface heat flow, they outgas only ~70% of $^3$He and only ~50% of radiogenically produced $^{40}$Ar. This suggests that Ar is not really 'missing'; it is still contained in the Earth due to the relative inefficiency of mantle outgassing. These results were confirmed by Tucker et al. (2022) who also showed that subducted atmosphere derived Ar could reasonably reproduce mantle $^{40}$Ar/$^{36}$Ar.

## 4.2 Distribution and Character of Noble Gases in the MORB-Source Mantle

The small variance in MORB $^3$He/$^4$He (7.94±0.99 R$_a$; Moreira, 2013) relative to that in OIB remains a defining difference between these two mantle domains, with data from earlier studies summarised in Barfod et al. (1999; see Figure 1a), Graham (2002), and Moreira (2013). There are significant departures from this range in ridge segments associated with plume components. These include the Reykjanes ridge in the North Atlantic (Hilton et al., 2000); the Shona and Discovery anomalies in the South Atlantic (Moreira et al., 1995); and the Saint Paul-Amsterdam plateau on the Southeast Indian Ridge (Graham et al., 2014). Correlation with radiogenic tracers, such as Sr, and ridge depth can be used to argue for ridge-entrained plume components (Moreira, 2013). A detailed study of 150 samples from 1500 km of the Southeast Indian ridge (Graham et al., 2014) shows $^3$He/$^4$He that varies from 7.5 to 10.2 $R_a$ with spectral analysis identifying wavelength power at 1000, 100, and 30 km. That $^3$He/$^4$He varies on this length scale with axial depth is interpreted to be due to coupling between regional variations in mantle temperature, melt production, and $^3$He/$^4$He controlled by folding and stretching of heterogeneities during regional (1000 km) and mesoscale (100 km) mantle flow, with the 30 km length scale likely imposed by the partial melting process.

A significant insight into mantle dynamics has been provided by developments in understanding neon provenance in different mantle sources. Samples from the mantle with a $^{20}$Ne/$^{22}$Ne isotopic component higher than air ($^{20}$Ne/$^{22}$Ne=9.80) were first unambiguously identified in gas rich MORB and OIB (Sarda et al., 1988) and Zaire cubic diamonds (Ozima and Zashu, 1988). Later work (Honda et al., 1991; 1993; Hiyagon et al., 1992; Poreda and Farley, 1992; Ballentine, 1997; D. Harrison et al., 1999; Caffee et al., 1999) confirmed the conclusion of Ozima and Zashu (1988) that this was a Solar-like isotopic signature, in both the deep and shallow mantle. Determination of whether this represented a trapped implanted Solar wind component in accretionary material (Ne-B ~12.5 ; Black, 1971; 1972a; 1972b; Moreira, 2013; Moreira and Charnoz, 2016; Péron et al., 2017) or a Solar Nebula value ($^{20}$Ne/$^{22}$Ne = 13.36 ± 0.09; Heber et al., 2012) has been crucial in defining the basis for interpreting the origin of inert gases in the mantle and identification of the processes that have variably preserved and altered their isotopic and element composition in the Earth's mantle (Porcelli and Ballentine, 2002; Moreira, 2013; Mukhopadhyay and Parai, 2019).

The source $^{20}$Ne/$^{22}$Ne value in mantle-derived samples is obscured by sample interaction with air and seawater on eruption. High precision data from gas rich MORB 'popping rocks' showed an air-mantle mixing line, with $^{20}$Ne/$^{22}$Ne plotted against $^{21}$Ne/$^{22}$Ne, appearing to stop at $^{20}$Ne/$^{22}$Ne ~ 12.5 (Moreira et al., 1998; see also Figure 6). Early evaluation of the plume-source mantle suggested a similar (Trieloff et al., 2000), but equally ambiguous value (D. Harrison et al., 1999; Ballentine et al., 2001). Mantle-derived carbon dioxide natural gases have provided a useful sample resource. Bravo Dome is a magmatic carbon dioxide gas field in Harding County, New Mexico. The natural addition of a crustal radiogenic 'spike' to the magmatic gases provided an unambiguous assessment of the upper mantle $^{20}$Ne/$^{22}$Ne (12.49 ± 0.04) corroborating the limit found in MORB (Ballentine et al., 2005; Ballentine and Holland 2008; Holland and Ballentine, 2006). In contrast, $^{20}$Ne/$^{22}$Ne values plume samples have subsequently been found higher than the shallow mantle (MORB-source) limit (D. Harrison et al., 1999; Yokochi and Marty 2004; Mukhopadhyay, 20126).

The convergence of the isotopic composition of Ar, Kr and Xe in well gases (Ballentine et al., 2005; Ballentine and Holland 2008; Holland and Ballentine, 2006; Holland et al., 2009) with those in gas rich Atlantic MORB (Moreira et al., 1998; Parai et al., 2019) indicates that the magmatic fluids in continental gases are similarly sourced. Residual gases in basalt glasses can undergo extensive elemental fractionation during magma outgassing. Continental magmatic gases could represent the complementary, but also elementally fractionated, component. In addition, the convergence of the underlined elemental composition of well gases with gas-rich MORB also shows that significant elemental fractionation has not occurred. This means that this sample resource can be used to determine the relative elemental abundance of noble gases in the mantle – confident that we have excluded eruptive processing or crustal contributions (Ballentine and Holland, 2008; Holland and Ballentine, 2006).

In addition to information about the accretionary source(s) and early-Earth environment, the heavy noble gases Ar, Kr, and Xe show an overprint in the mantle component (distinct from the eruptive addition of air) that has an elemental and isotopic composition identical to that of dissolved air in seawater/marine pore fluids. This overprint accounts for all of the well gas mantle $^{36}$Ar and 80% of the $^{130}$Xe, isotopes little affected by radiogenic sources in the mantle and is interpreted as a clear signature of volatile element recycling into the mantle (Holland and Ballentine 2006; Sumino et al., 2010; Kobayashi et al., 2017). This is in contrast to earlier work which viewed subduction as an efficient barrier to volatile subduction (Staudacher and Allègre, 1988).

Noble gases have a high solubility in amphibole and ring-structure bearing minerals (C.R.M. Jackson et al., 2013; 2015), providing a transport mechanism in cold subduction zones (Smye et al., 2017). Observational evidence for the unfractionated subduction of a seawater or marine pore fluid signature has been preserved in hydrated mineral inclusions that have been exhumed from >100 km depth from multiple subduction zones (Sumino et al., 2010; Kobayashi et al., 2017). Subduction plays a central role in transport of seawater noble gas signatures into the mantle (Holland and Ballentine 2006; Ballentine and Holland, 2008; Sumino et al., 2010; Kendrick et al., 2011; 2013; Chavrit et al., 2016; Kobayashi et al., 2017; Smye et al., 2017), with Xe isotope systematics being used to show that 80–90% of the Xe in both the MORB and OIB-source mantle is recycled atmosphere-derived Xe (Parai and Mukhopadhyay, 2015; Tucker et al., 2012).

The broader impact of seawater noble gas subduction can be seen in samples from the Southwest Indian Ridge (Parai et al., 2012). This study is one of the first systematic applications of Ne isotopes to correct for eruption related air-contamination. Fourteen samples from across 1100 km of ridge show correlations between $CO_2/^3$He and Ne, Ar and Xe isotopic compositions as a function of longitude. The total amplitude in the resolved mantle $^{40}$Ar/$^{36}$Ar is between 12,000 to 50,000. Using $^3$He/$^4$He to minimise any plume influence, significant heterogeneity in Ar and Xe isotopes remains. This is a variance in isotopic character that is likely controlled by heterogenous recycling or mixing of volatiles into the ridge mantle source which, similar to the $^3$He/$^4$He length scale variance (Graham et al., 2014), provides an observational constraint for mantle mixing models.

## 4.3 Distribution and Character of Noble Gases in the Mantle

The resolved MORB-source Ne isotopic composition ($^{20}$Ne/$^{22}$Ne~12.5) is indistinguishable from that of Solar wind irradiated meteoritic accretionary material (Black, 1972a; 1972b; Ballentine et al., 2005; Moreira, 2013; Péron et al., 2017; 2018). This source interpretation is consistent with identification of an upper mantle chondritic signature in both the Kr and Xe isotopic compositions (Holland et al., 2009; Caracausi et al., 2016). In contrast, the $^{20}$Ne/$^{22}$Ne of plumes or plume-influenced MORBs show maximum $^{20}$Ne/$^{22}$Ne values higher than MORB (Sarda et al., 2000; Yokochi and Marty 2004; Mukhopadhyay, 2012; Péron et al., 2016). These have been interpreted as a possible Solar Nebula gas signal preserved in the deep mantle (Solar $^{20}$Ne/$^{22}$Ne=13.36 ± 0.09: Heber et al., 2012) which may represent a discrete volatile component in the OIB mantle (Mukhopadhyay, 2012; Parai et al., 2019; Parai and Mukhopadhyay, 2021). Xenon isotopes from Iceland (Mukhopadhyay, 2012) provide evidence that the OIB-source mantle preserves a long-lived geochemical record, consistent with preservation of a Solar Nebula Ne signal. The xenon isotope ratios generated by the extinct radionuclides $^{129}$I and $^{244}$Pu, and therefore signals generated within the first circa 100 Myr, show that the OIB-source had I/Pu ratios lower than that recorded in the MORB-source mantle. This signal has also been used to argue for the preservation of an early volatile poor (low I/Pu; dry) accretionary stage in the deep mantle (Mukhopadhyay, 2012; Mukhopadhyay and Parai, 2019; Parai and Mukhopadhyay, 2021). The $^3$He/$^{22}$Ne elemental ratio in OIB is also similar to Solar Nebula values and distinct from the ~3× higher value observed in MORB. The difference has been attributed to magma ocean-atmosphere equilibration and punctuated degassing & early atmosphere loss (Honda and Macdougall, 1998; Coltice et al., 2011; Moreira, 2013; Tucker and Mukhopadhyay, 2014). It is important to note that while both MORB and OIB show evidence for having recycled noble gases in common, there are multiple lines of evidence now that show they preserve a different history of planetary volatile accretion, radiogenic ingrowth, and mixing within the mantle. This record is incompatible with models requiring the OIB-source mantle to provide the volatile source, or flux, for the MORB-source mantle.

Our conceptual understanding that underpins much of the modeling has focussed on a potential deep source for the excess $^3$He, and related heavier accretionary noble gases. This is needed to balance the observed $^3$He/$^4$He in MORB with the inferred concentration of $^3$He and radiogenic ingrowth from U+Th, now within a whole mantle convection setting (e.g., Porcelli and Elliott, 2008). The core has received considerable attention since it is large, formed early and has been isolated from the mantle for most of Earth history. While the earliest work suggested that helium would not partition into the forming core (Matsuda et al., 1993), updated silicate/metal partitioning conditions brought this back into consideration (Porcelli and Halliday 2001; Bouhifd et al., 2013). Most recently ab-initio thermodynamics have been used to derive the partition coefficients of noble gases between liquid iron and silicate melt at core-forming and silicate equilibration conditions (from 3500 K at 50 GPa to 4200 K at 135 GPa; Y. Li et al., 2022). While able to predict a concentration of helium in the core that could sustain the $^3$He in the silicate mantle, this study found the partition coefficients for Ne to be three orders of magnitude lower than He. This predicts $^3$He/$^{22}$Ne in the core to be three orders of magnitude higher than that observed in plume derived melts, which have near Solar Nebula values, and suggests that a core-source able to simply account for both $^3$He as well as related accretionary noble gases is unlikely.

A source for the high $^3$He concentrations and associated heavy noble gases in a stable silicate reservoir within the deep mantle have also been proposed. These have in part been inspired by heterogeneities at the core-mantle boundary that are seismically observed (LLSVPs and ULVZs) and appear to be related to plumes (French and Romanowicz, 2015; Williams et al., 2019; Heyn et al., 2020) and their likely connection to recycled oceanic crust (Hofmann and White, 1982; Christensen and Hofmann, 1994; Kendall and Silver, 1996; Coltice and Ricard, 1999; Sobolev et al., 2000; Tolstikhin et al., 2006; Thompson et al., 2019; Gréaux et al., 2019; M.G. Jackson et al., 2021). This link is strengthened by geodynamical models that show that early oceanic crust subduction can sequester $^3$He and other primordial components in ancient LLSVP precursors (Jones et al., 2021). While not always explicit in the publications that explore the $^3$He origins, accretionary models can be divided into those that would produce a Solar wind irradiated signature in both deep silicate reservoir and MORB-source mantle – and those that would preserve a Solar Nebula signature in the deep silicate reservoir (OIB-source) but would need the residual mantle to be overprinted with a Solar wind irradiated signature that evolves to source MORB.

The subduction of early and dry oceanic crust (to facilitate volatile subduction efficiency) containing a high load of Solar wind irradiated carbonaceous chondritic dust would result in adding a Solar wind irradiated noble gas signature to the mantle, concentrating this signature at the core-mantle boundary (Tolstikhin and Hofmann, 2005). Continuous subduction of cosmogenic material over Earth history (Allègre et al., 1993; Anderson 1993) does not simply account for Solar-like noble gas concentrations in the mantle (Stuart, 1994). In contrast, subduction does provide a mechanism for the occurrence of marine pore fluid signatures in both MORB and OIB (Holland and Ballentine, 2006; Mukhopadhyay, 2012). Also, ingassing of nebular gas from a gravitationally accreted early atmosphere into a magma ocean (Mizuno et al., 1980; Harper and Jacobsen 1996; Porcelli et al., 2001; Yokochi and Marty, 2004; Mukhopadhyay, 2012; Tucker and Mukhopadhyay, 2014) would add a Solar Nebula signature into the mantle. Incomplete degassing of the deeper mantle and isolation at depth early in Earth history (Wen et al., 2001), or foundering of dense undegassed melts (Lee et al., 2010) to form D'' both provide a mechanism that could account for the Solar Nebula signal seen in some OIBs.

Mechanisms not associated with the deep seismic features that can produce $^3$He/$^4$He heterogeneity more broadly dispersed in the mantle may involve high viscosity regions in the mantle which are predicted to escape efficient mixing (Manga, 1996; Merveilleux du Vignaux and Fleitout, 2001; Manga, 2010; Ballmer et al., 2017). This consistent with statistical analysis of radiogenic isotope systems that original identified a common component (the aforementioned PREMA, FOZO, or 'C') categorising this as stochastically distributed small-scale isotopic heterogeneities (Stracke et al., 2022).

The redetermination of the reference chondritic halogen (Cl, Br and I) concentrations has been made possible by the application of neutron irradiation noble gas mass spectrometry (NI-NGMS), an analytical technique successfully employed for decades in Ar-Ar geochronology and related systems (Burgess et al., 2002; 2009; Ruzié-Hamilton et al., 2016). Contrary to previous bulk sample analysis, this enabled mg quantities of sample to be picked from the least altered and contaminated portions of the meteorite samples (Clay et al., 2017). Prior studies provided significantly higher halogen concentrations in chondrites and required exceptional accretionary processes to account for them (e.g., Dreibus et al., 1979; Lodders, 2003). Authors still contend that the bulk analysis of chondrite samples produces a more representative analysis (Lodders and Fegley, 2023) but rely on speculation where they cite '...suspect analytical problems' with NI-NGMS. Speculation and suspicion notwithstanding, the Clay et al. (2017) results provide for a monotonous behavior of the halogens during accretion, depleted in the Bulk Silicate Earth in proportion to their condensation temperature, similar to other lithophile elements. In contrast to their BSE abundance, 80 to 90% of the BSE halogens are concentrated in the oceans and continental crust (Burgess et al., 2002), cannot have been emplaced by accretion after core formation (too low concentration in chondrites), and are likely to have been efficiently extracted from the accreting Earth in water-containing melts since they are highly hydrophilic. This is consistent with other evidence for a volatile or water-rich main accretionary phase (Clay et al., 2017).

Extra-terrestrial and magmatic are susceptible to acquiring an overprint from the high halogen concentrations found at Earth's surface. The processes controlling an overprint of magmatic signatures were investigated in detail by Broadley et al. (2017) to develop protocols that minimised this analytical complexity. This widened the samples available to study and contributed to showing significant recycling of the terrestrial surface halogens to the subcontinental lithospheric mantle (Sumino et al., 2010; Broadley et al., 2016; Kobayashi et al., 2019). Indeed, sudden release of subducted and accumulated lithospheric mantle halogens at the end of the Permian, during the Siberian Flood Basalt eruptions, may have significantly contributed to the surface environmental degradation that resulted in one of the largest extinctions in our planetary record (Broadley et al., 2018). Consideration of how surface halogens are added to the oceanic crust shows them to be dominantly mineral bound with Cl concentrations in hydrated oceanic crust at concentrations where subduction is readily able to account for the Cl in the MORB source mantle (Barnes and Sharp, 2006; John et al., 2011; Kendrick et al., 2011; 2013; 2017; Beaudoin et al., 2022), pointing to a significant overprint of surface halogens over any accretionary signals in mantle halogens today. The subduction process preserves the relative abundance of the different halogens, enriching them compared with non-mineral bound volatile species such as the noble gases in both the MORB-source shallow mantle (Kobayashi et al., 2017; 2019; Bekaert et al., 2021) and Hawaiian plume source (Broadley et al., 2019).

## Discussion and Conclusions

In the review above we have tried to highlight areas of progress in our understanding of the evolution of Earth's crust and mantle with a particular emphasis on oceanic crust recycling, constraints from noble gas systematics, and use of novel geochemical systems. Of course, new research also unearths new conflicts, contradictions, and areas where we clearly lack an understanding of how the Earth's mantle evolved. We end this chapter by highlighting a few important areas where we still lack sufficient understanding.

1. Why are HIMU OIBs so rare if they represent recycled oceanic crust over Earth history (Chauvel et al., 1992; Stracke et al., 2005)? Indeed, if formation and subduction of oceanic crust occurred over most of Earth history, large quantities of oceanic crust should have been reinjected into the mantle (see section 2.1). The rarity of HIMU OIBs might be explained by the necessary combination of two factors: an old age (> 1.5 Ga) and the complete absence of a sedimentary layer that would immediately hide the geochemical characteristics of pure oceanic crust (see Chauvel et al., 1992 or Shimoda and Kogiso, 2019). If oceanic crust is recycled for less than about 500 Ma, it would not have time to develop the radiogenic Pb isotopic compositions that characterise HIMU OIB but it could potentially be traced using trace element characteristics such as elevated Ce/Pb or Nb/La ratios. The geochemical community has maybe not paid enough attention to the message carried by trace elements and too much to the fingerprints given by isotopic systems.

2. Using trace elements and Nd isotopes mass balance calculations, Hofmann et al. (2022) demonstrated that the chemical budget between present-day mantle (as sampled by both MORB and OIB) and continental crust requires that they both originate from a 'residual mantle' formed early on. The enriched reservoir complementary to this 'residual mantle' is either totally hidden and never sampled by volcanism or it has been lost to space. Further work combining geochemistry and geodynamics is required to try to understand where it may have been preserved and whether it could be contained in the LLSVPs (e.g., Jones et al., 2021). Alternatively, maybe it is sampled by kimberlites as suggested by Woodhead et al. (2019). Further geochemical studies on kimberlites, that also suggest that PREMA is preserved in the LLSVPs (Guiliani et al., 2021) might help provide an answer.

3. The negative correlation between He and W isotopic compositions as observed by Mundl et al. (2017) and Mundl-Petermeier et al. (2019; 2020) among OIBs raises puzzling questions for its origin. Indeed, with helium being a noble gas and tungsten a siderophile element, no simple melting or fractionation process can explain such correlation. Mundl-Petermeier et al. (2020) suggest that mixing of three distinct reservoirs could account for the observed correlation. They locate one of these reservoirs in the general convective mantle. The other two are the LLSVPs and ULVZs at the core-mantle boundary that are assumed to be isolated for billions of years from the rest of the mantle. If pristine Hadean material were to be stored in the deep Earth it should potentially also be recognizable in terms of [142]Nd isotopic compositions. However, it is not the case since at this stage since all analysed OIBs except maybe for one (Peters et al., 2018) do not show any significant deviation from the [142]Nd BSE value. Alternative explanations remain to be found. For example, very recently, Peters et al.

(2023) suggested that interaction between metal and silicate occurring at low pressures during core formation could potentially account for a reservoir with very negative tungsten isotopic composition but it remains that one cannot completely eliminate the possibility that the He-W correlation might be fortuitous with helium and tungsten tracing independent sources that only mix when melting proceeds. Constraints from dynamical modeling on this question can be found in Ferrick and Korenaga (2023).

4. It has become clear that multiple sources and accretionary processes deliver noble gases and related volatiles to the mantle together with a mechanism or mechanisms that variably preserve their distinct signatures in the OIB and MORB-source mantle. In the broadest sense we need to build a mantle that preserves and traps Solar Nebula gases, exemplified by $^3$He, in the OIB-source, with the conjunction of stabilised silicate material at the core mantle boundary providing a preservation mechanism for this component in the mantle. While this component provides its high $^3$He/$^4$He, Nebular $^{20}$Ne/$^{22}$Ne and $^3$He/$^{22}$Ne, this does not and cannot significantly contribute to the mantle that supplies MORB (Mukhopadhyay, 2012). Delivery of a Solar wind irradiated Ne component together with chondritic Kr and Xe must overprint any Solar Nebula residue. There is still considerable infall of accreting material after the magma ocean phase of planetary formation and the opportunity for dry subduction of Solar wind irradiated chondritic material – but the need for a $^3$He source or flux remains, unless we have the concentration of $^3$He in the MORB-mantle wrong (Ballentine et al., 2002).

5. We have taken here the viewpoint that oceanic lithosphere with dense crust is the prime source for deep mantle heterogeneity. We have ignored here the potential interaction of the oceanic lithosphere with the mantle transition zone that is between the 410 km and 660 km seismological discontinuities. In particular the endothermic nature of the 660 km discontinuity has been used to argue for the potential of slabs to slow down or stagnate in the transition zone (Christensen, 1996) as is observed for example in the western Pacific (Fukao et al., 2006). In addition, while it has long been argued that subducted oceanic crust is denser than ambient upper mantle (e.g., Ringwood, 1967) it is less dense at the top of the lower mantle (Irifune and Ringwood, 1993) which could allow for further accumulation of oceanic crust. Trapping of the oceanic crust at this depth could occur as it is subducted into the lower mantle helped by rheological contrasts (e.g., van Keken et al., 1996) or due to separation of stagnant slab segments (e.g., Feng et al., 2021). It could also occur over time as the subducted oceanic crust is entrained by upwellings and then retained below the 660 km discontinuity (van Summeren et al., 2009; Nakagawa et al., 2012; Yan et al., 2020). It also has been argued that slabs stagnate a little deeper in the lower mantle (Ballmer et al., 2015) potentially leading to further reduction in the efficiency of oceanic crust subduction. Seismic observations provide direct evidence for heterogeneity in the mantle transition zone (Goes et al., 2022; Yu et al., 2023) that suggests the transition zone modulates convection. Since the seismic observations are best explained by a mechanical mixture of basalt and refractory peridotite (Goes et al., 2022), it is unlikely that the transition zone completely blocks descent of the oceanic lithosphere to the deep mantle, as the transition zone (with a volume of 10% of the Earth's mantle) would become completely filled by oceanic crust if subduction only proceeds to the base of the transition zone. The slower mixing predicted by these observations may help explain why the models presented in Brandenburg et al. (2018), that while providing reasonable global overlap with observations of HIMU, EM1, and the MORB-source, effectively predict an upside down mantle with a depleted lower mantle and enriched upper mantle (Tucker et al., 2020).

Aside from future progress in the individual scientific approach we expect that significant progress can be made by interdisciplinary research that specifically combines geochemical and geophysical observations with geodynamical modeling that incorporate experimental and theoretical constraints on the constitution of Earth's mantle, which would form an expansion of the work presented by, for example, van Keken and Ballentine (1998, 1999), Xie and Tackley (2004a; 2004b) and Brandenburg et al. (2008).

## Acknowledgements

This work was financially supported by the ERC advanced grant SHRED awarded to CC (grant agreement n°833632). CJB thanks the Carnegie Institution for Science for a Tuve Fellowship that allowed for discussions and writing parts of the manuscript.

## Figures

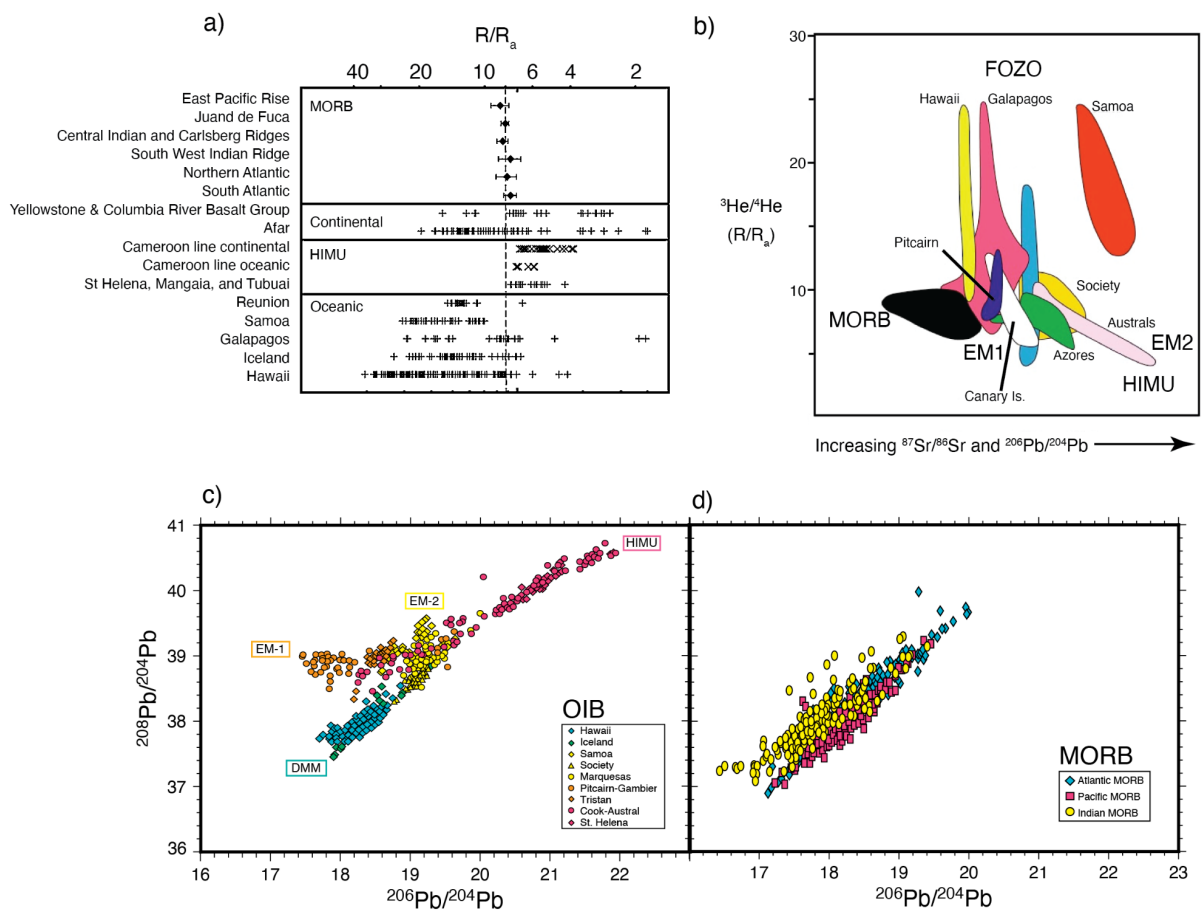

Figure 1. a) Redrawn from Barfod et al. (1999): helium isotopic ratios suggest a dramatic more homogeneous upper mantle (assuming MORBs sample this) than the lower mantle (assuming OIBs sample this); R=$^3$He/$^4$He, R$_a$=R of the present day atmosphere. b) Redrawn from van Keken et al. (2002): Data envelopes for various hotspots suggest mixing lines between MORB, EM1, EM2, HIMU, and FOZO endmembers; c) Redrawn from Hofmann (2014): heterogeneity shown in Pb-Pb isotopes in OIBs with apparent mixing lines between endmembers. d) Redrawn from Hofmann (2014) on the same scale as panel c for MORB. While the spread of isotopic ratios in MORBs is somewhat less than that in OIBs the MORB source is clearly not nearly as homogeneous as is suggested from the helium isotopic ratios (panel a).

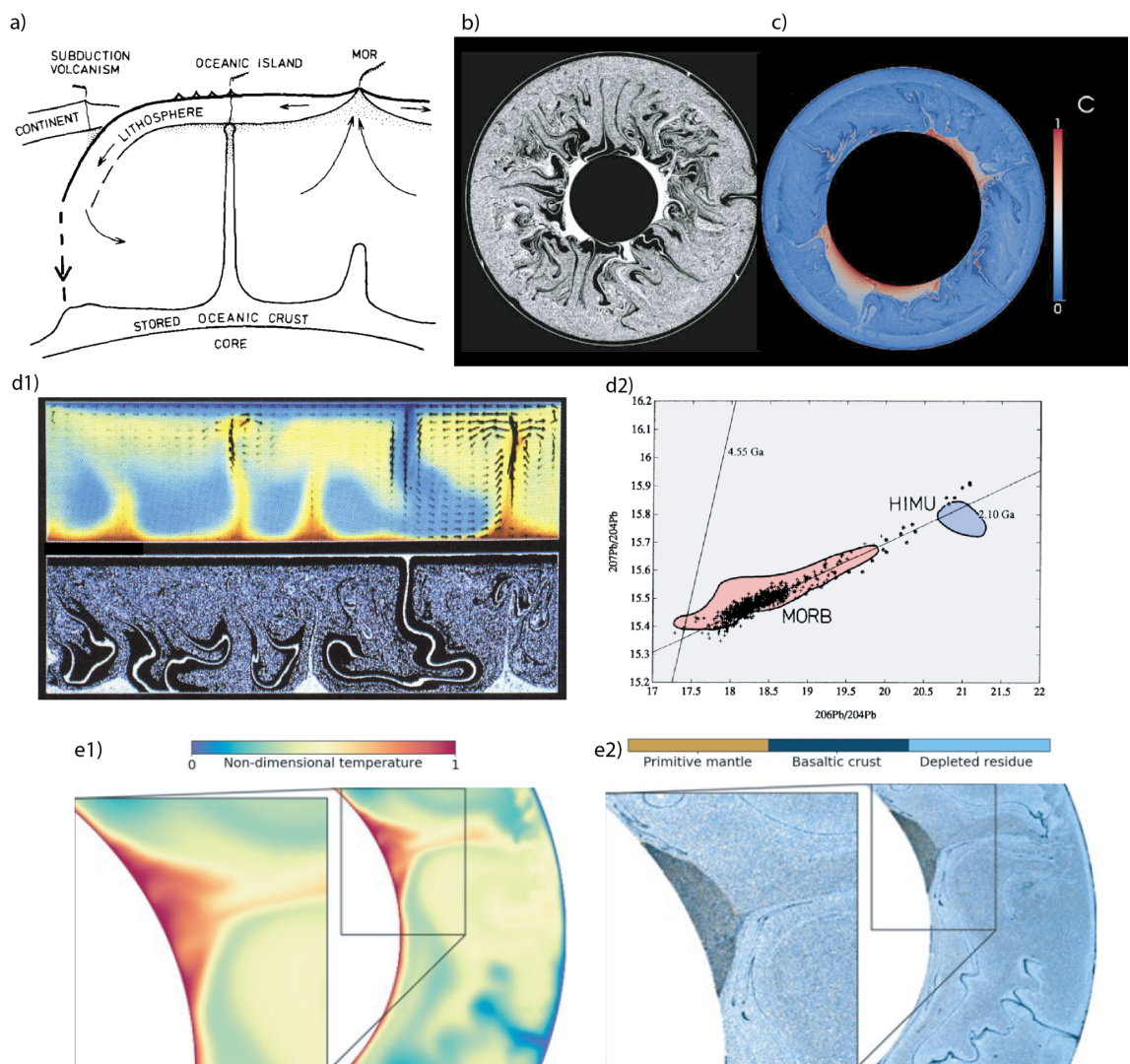

Figure 2. a) Cartoon from HW82 providing the suggestion for the importance of oceanic lithosphere recycling and storage of oceanic crust at the base of the mantle for OIB chemistry. b) From Brandenburg et al. (2008). Eclogite tracers are shown in white indicate the concentration of oceanic crust at the base of the mantle (model with basalt excess density of 3.6% with respect to ambient mantle); c) From Nakagawa et al. (2012). Model computed with self-consistent material properties showing accumulation of oceanic crust in dense piles (indicated by high C) at the base of the mantle and, to some extent, at the base of the transition zone. d1+d2) from Christensen and Hofmann (1994). d1: Ocean crust segregation and recycling modeled with kinematic plates; top frame shows nondimensional temperature and lower frame shows eclogite component in white. d2: Pb-Pb isotopes derived from the model reproducing approximately the MORB and HIMU fields (but see argument about transit time scaling in the paper). e1+e2) from Jones et al. (2021). e1: temperature showing high temperature in the thermochemical pile with convective motions. e2: tracers showing accumulation of oceanic crust at the base of the mantle. In this application it is suggested that the thermochemical piles can trap primitive mantle potentially providing a natural explanation for the preservation of ancient heterogeneity.

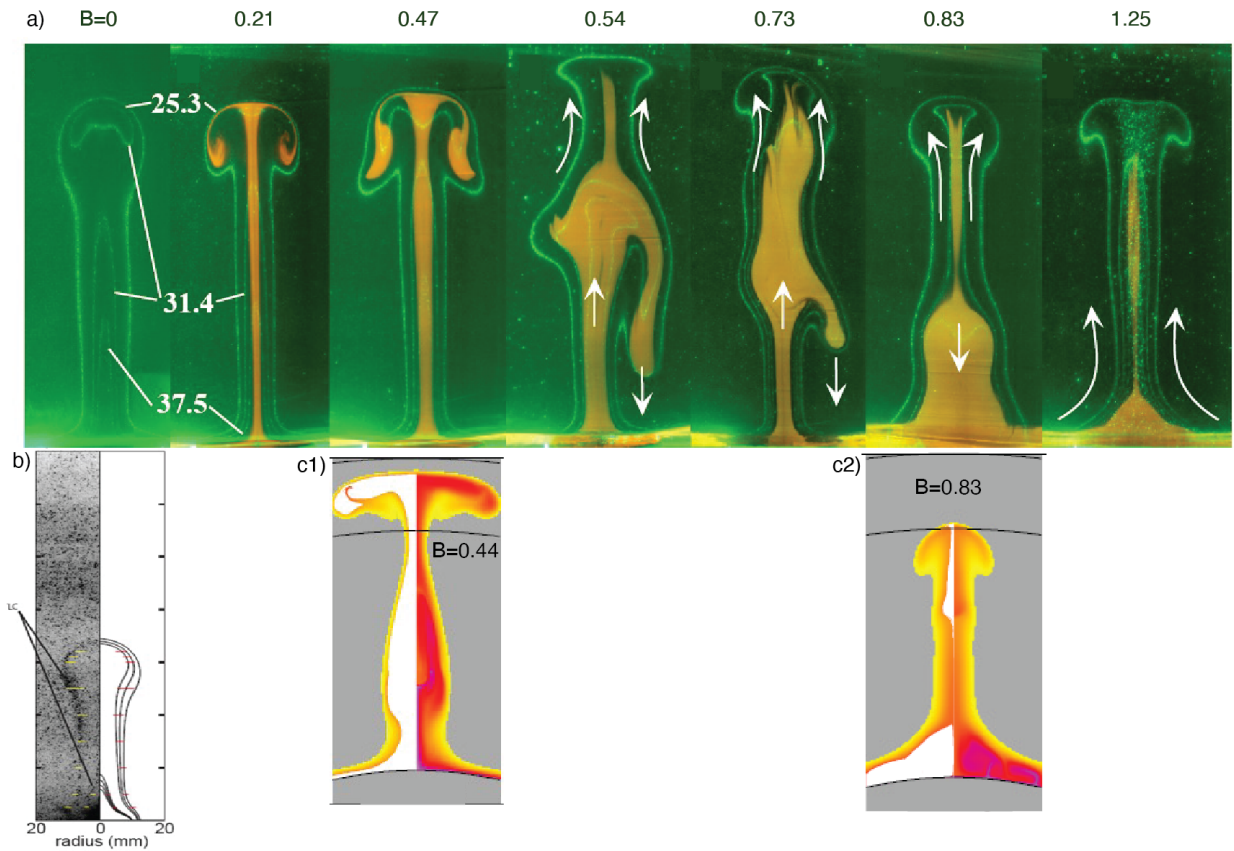

Figure 3. Examples of entrainment of a dense layer by thermochemical plumes. a) From Kumagai et al. (2008). Laboratory plumes starting from a heater entrain a denser orange layer. The green color is from the laser sheet that illuminates the near-cylindrical plume. Contour lines labeled in the first two frames on the left indicate temperatures in the fluid, a technique made possible by the use of crystals that are opaque under a narrow temperature range. From left to right the density of the orange fluid increases leading to increasing less efficient entrainment by the thermal plume. B is the non-dimensional buoyancy number that indicates the ratio of compositional to thermal buoyancy. b) From Vatteville et al. (2009): comparison of a starting plume between a similar laboratory set up as that in a) but without dense layer. On the left the opaqueness indicates the ranges where the temperature is between a certain interval. Horizontal lines indicate the same temperature ranges from laser diffraction. On the right is the numerical simulation of this experimental setup showing excellent agreement. c) Numerical simulations from Lin and van Keken (2006b) with colors indicate temperature. The white layer is dense material with a buoyancy number indicated. Comparison with the results in frame a) suggest similar dynamics is suggested between laboratory and numerical experiments, showing the complementarity of these two approaches.

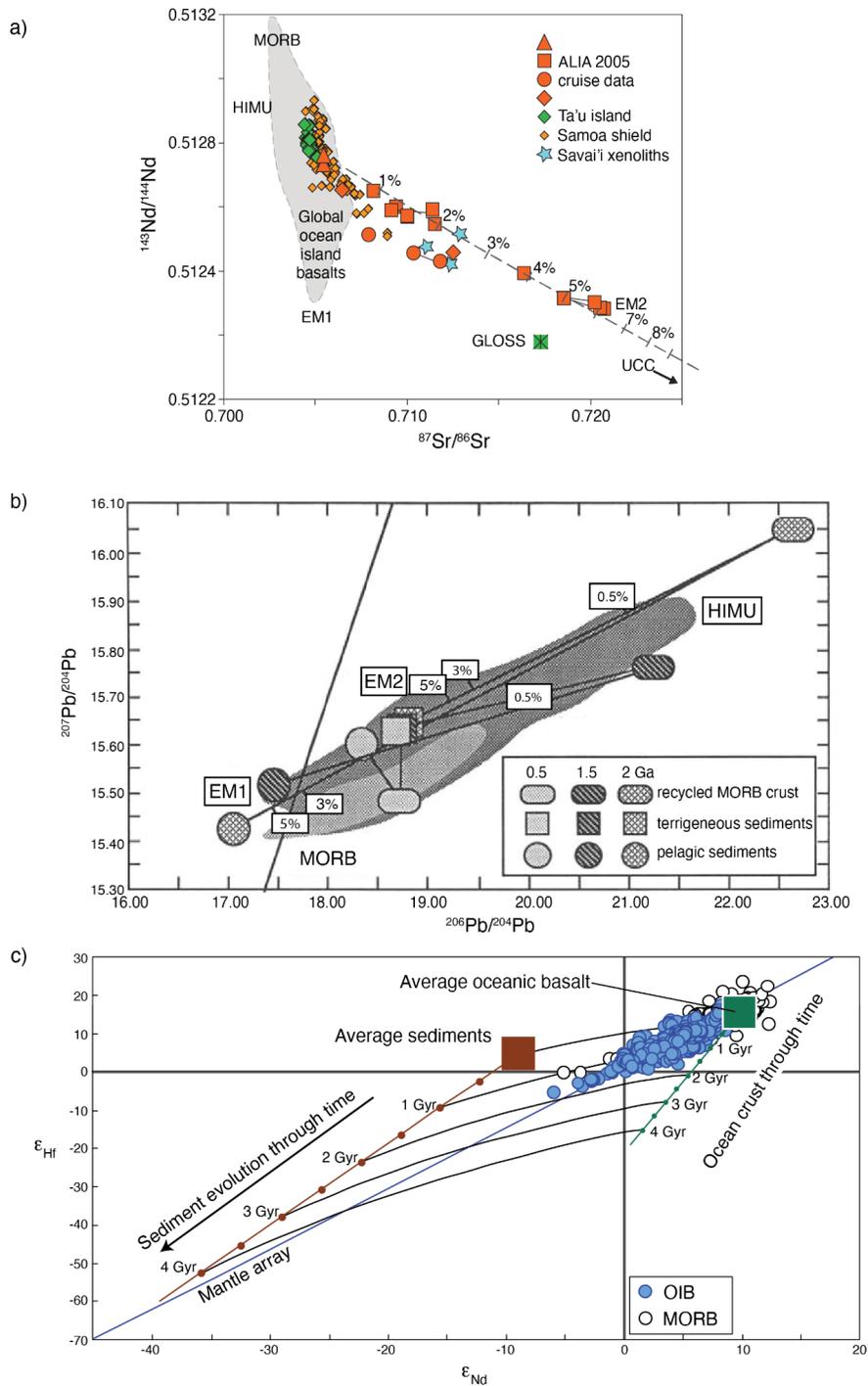

Figure 4. Specific recycling of oceanic crust and associated sediments and how it can be traced using radiogenic isotopic systems. a) In 2007, M.G. Jackson et al. reported extremely radiogenic Sr isotopes for a basaltic sample collected along the Samoan archipelago in the Pacific. These extreme values resemble what is known on continental crust or in oceanic sedimentary material (GLOSS). Such high Sr isotopic compositions can only be explained by the recycling of crustal material in the Earth's mantle. b) the range of Pb isotopic compositions

of ocean island basalts (dark grey field) and MORB (light grey field) can be reproduced by the involvement of recycled basaltic crust and associated sediments with varying ages (Chauvel et al., 1992); HIMU compositions lie next to pure oceanic crust recycled more than 1.5 Ga ago, while EM1 and EM2 compositions correspond respectively to old pelagic and terrigenous sediments; c) the Hf-Nd mantle array requires the involvement of both recycled oceanic crust and oceanic sediments in the source of OIB and MORB (Chauvel et al., 2008).

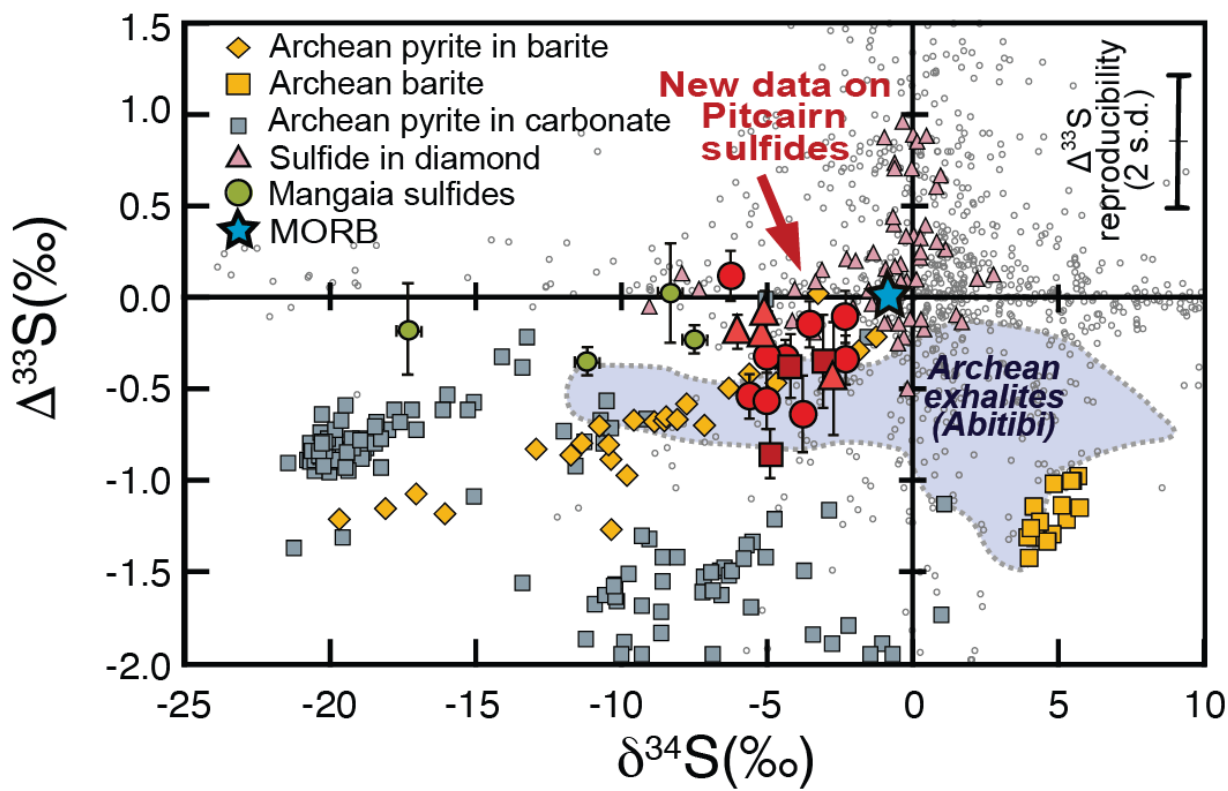

Figure 5: Sulphur isotopes diagram modified from Delavault et al. (2016) showing the existence of mass-independent fractionation anomalies ($\Delta^{33}$S different from 0) in sulfides from Mangaia island (Cabral et al., 2013) and Pitcairn seamounts (Delavault et al., 2016).

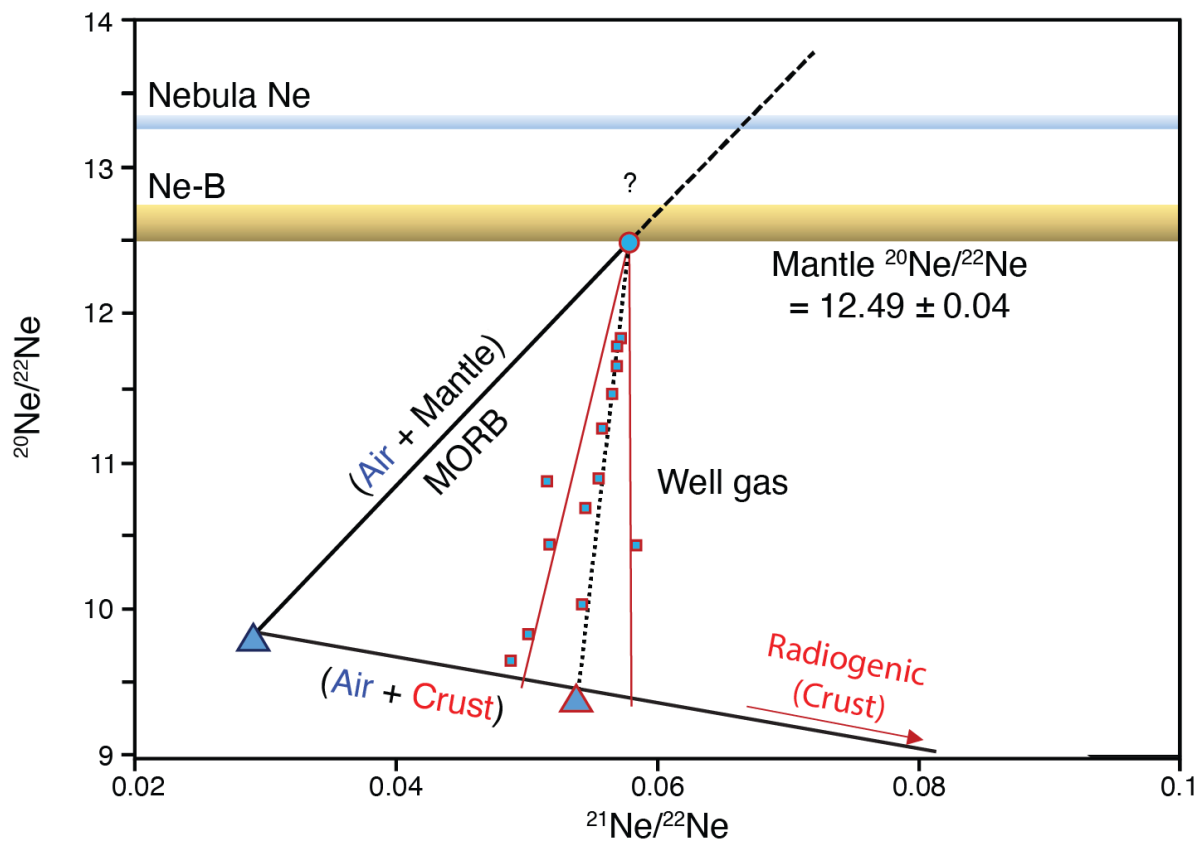

Figure 6. The three isotopes of neon provide a clear view on the origin of Ne in the Earth's upper mantle. Air is ubiquitous in all samples measured. Early work was not able to unambiguously distinguish the mantle source composition and resolve between Ne isotopic compositions related to a Solar wind irradiated accretionary source (Ne-B) vs Solar Nebula gases with a higher $^{20}Ne/^{22}Ne$ ratio which require fundamentally different processes to be operating in the planetary accretion of volatile elements. Ancient groundwater in the continental crust contains dissolved air with a crustal radiogenic component. When magmatic gases in the continental contact this fluid it generates a unique mixing line in the neon system. The intersection or triangulation of this mixing line with the MORB-air mixing line unambiguously determines the mantle source Ne isotope composition. The mantle Ne isotopic value determined is indistinguishable from Solar wind irradiated accretionary material (Ne-B)(Ballentine et al., 2005; 2008; Holland and Ballentine, 2006).